\documentclass[12pt]{article}
\usepackage{graphicx}

\newcommand\pubdate{\today}

\def\Title#1{\begin{center} {\Large #1 } \end{center}}
\def\Author#1{\begin{center}{ \sc #1} \end{center}}
\def\Address#1{\begin{center}{ \it #1} \end{center}}

\newcommand\pubblock{\rightline{\begin{tabular}{l}  \\ 
         \pubdate  \end{tabular}}}
\newenvironment{Abstract}{\begin{quotation}  }{\end{quotation}}
\newenvironment{Presented}{\begin{quotation} \begin{center} 
             PRESENTED AT\end{center}\bigskip 
      \begin{center}\begin{large}}{\end{large}\end{center} \end{quotation}}

\begin{document}
\begin{titlepage}
 \pubblock
\vfill
\Title{Recent results from the CMS Proton Precision Spectrometer}
\vfill
\Author{Christophe Royon}
\Address{Department of Physics and Astronomy, The University of Kansas, Lawrence, USA}
\vfill
\begin{Abstract}
The Precision Proton Spectrometer (PPS) is a new subdetector of CMS that provides a powerful tool for the advancement of beyond standard model searches. We present recent results obtained with the PPS subdetector illustrating the unique sensitivity achieved using proton tagging.
\end{Abstract}
\vfill
\begin{Presented}
DIS2023: XXX International Workshop on Deep-Inelastic Scattering and
Related Subjects, \\
Michigan State University, USA, 27-31 March 2023 \\
     \includegraphics[width=9cm]{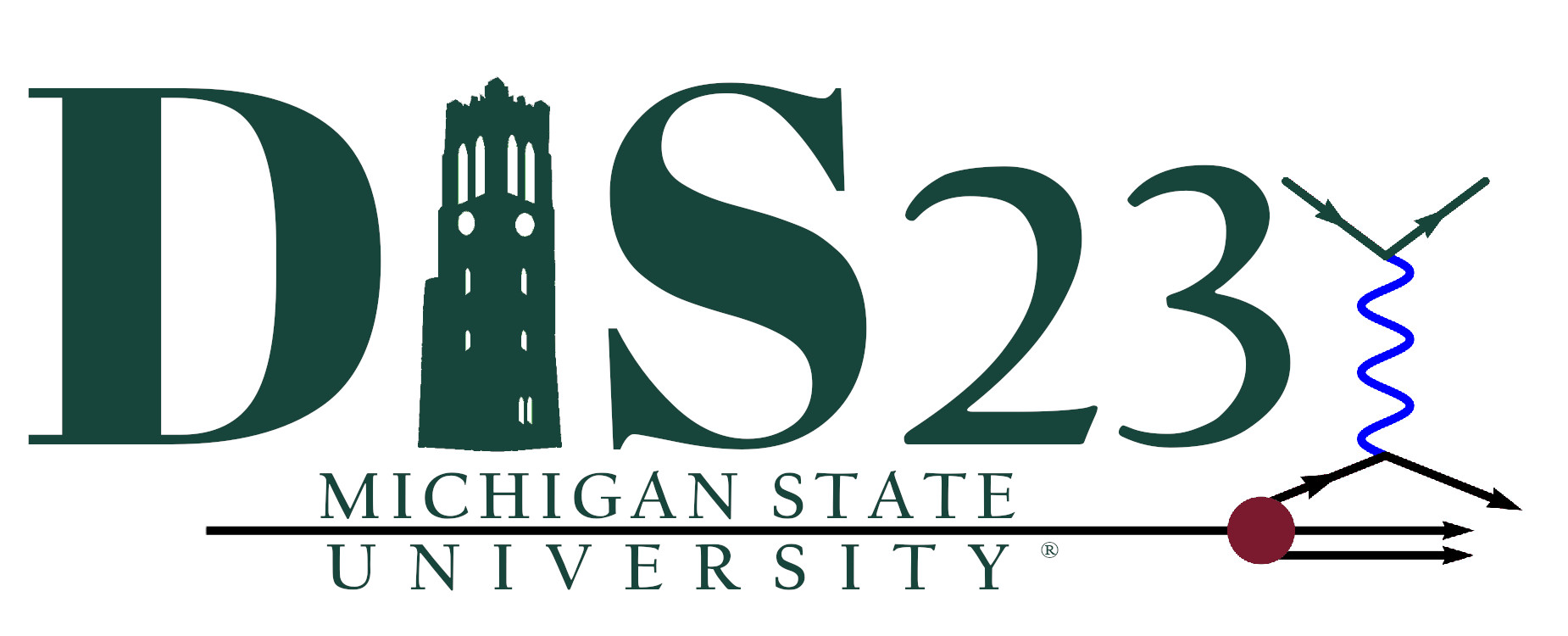}
\end{Presented}
\vfill
\end{titlepage}

\section{Introduction: photon induced processes}

\begin{figure}
\centering
\includegraphics[width=0.8\textwidth]{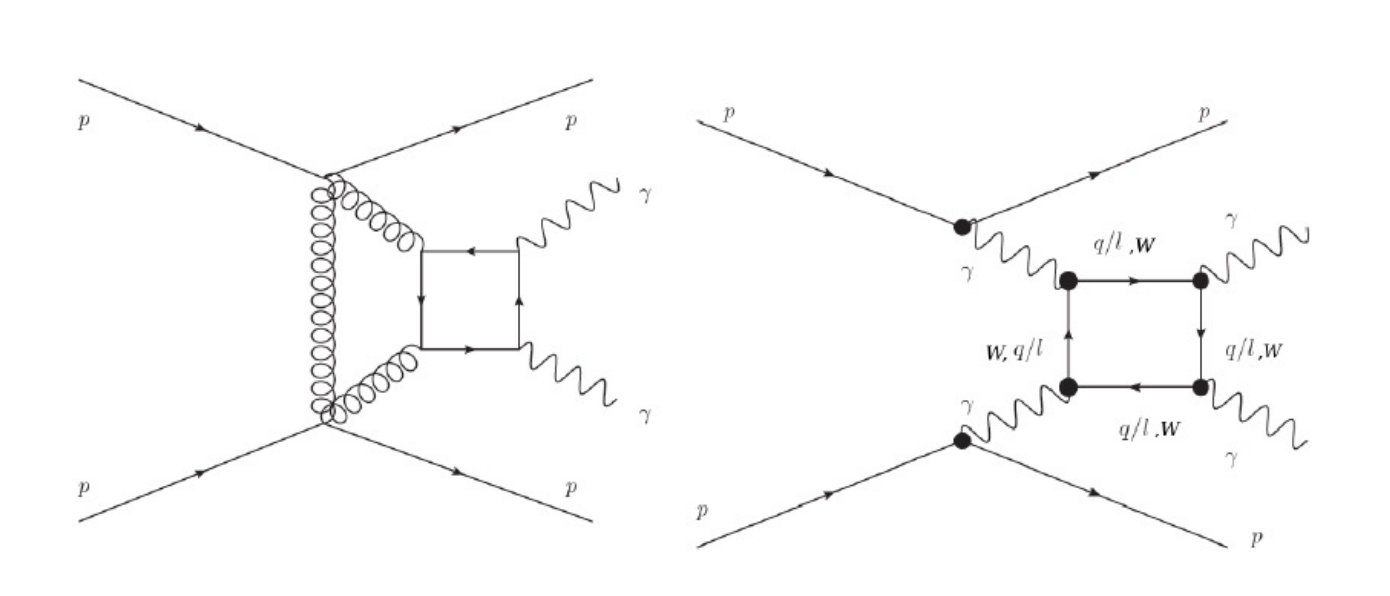}
\caption{Exclusive diphoton production. Left: QCD diagram via gluon exchange. Right: QED diagram via photon exchange.}
\label{fig1}
\end{figure}

We discuss recent results from the CMS and TOTEM collaborations using the Precision Proton Spectrometer (PPS). PPS allows detecting and measuring intact protons in $pp$ collisions at the LHC. 
In standard runs at the LHC, PPS has a good acceptance in diffractive  masses typically between 450 and 1400 GeV when both protons are detected after collisions~\cite{tdr}.  This shows the possibility to look for beyond standard model (BSM) event production at the LHC using a completely new method. More than 100 fb$^{-1}$ of data was collected in Run II by the CMS and TOTEM collaboration.

As an example, let us consider the exclusive diphoton production as shown in Fig.~\ref{fig1}. The left diagram displays the QCD process via two gluon exchange while the right diagram displays photon exchanges. At LHC energies, in the acceptance of PPS above 450 GeV, photon exchanges completely dominate gluon ones by several orders of magnitude~\cite{gammagamma5}. If we measure two intact protons and two photons in CMS, we are sure that this is
a photon exchange. The same conclusion applies for exclusive productions of $WW$, $ZZ$, $t \bar{t}$, $\gamma Z$ which we will describe in turn. The LHC can be considered as a $\gamma \gamma$ collider.
 
\begin{figure}
\centering
\includegraphics[width=0.8\textwidth]{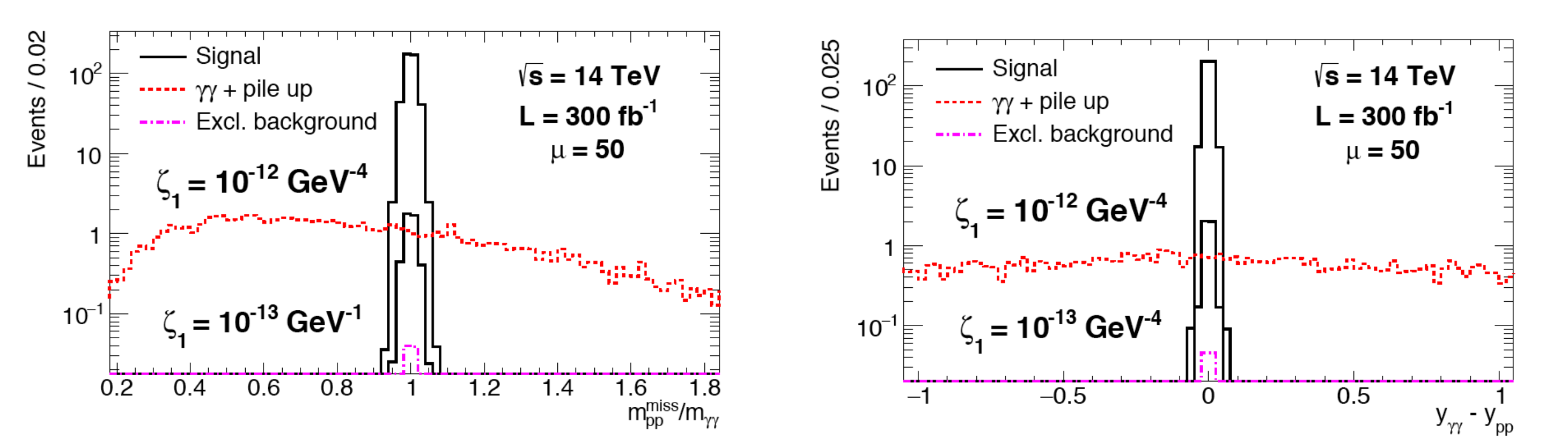}
\caption{Ratio between the diproton and diphoton mass and difference between  the diproton and diphoton rapidity for signal in black and pile up background in red.}
\label{fig2}
\end{figure}

Let us also discuss the advantages of detecting the intact protons in the final state using again the example of exclusive diphoton production. For our signal events, we detect all particles after interactions, namely the two intact protons and the two photons. The conservation of momentum and energy ensures that the mass and rapidity of the diphoton and the diproton systems are the same for signal. The leading background is due to pile up events where diphotons and intact protons originate from different proton-proton interactions (we recall that up to 50 interactions occur per bunch crossing at the LHC in standard running). The ratio of the diphoton and diproton masses as well as the difference in rapidity are shown in Fig.~\ref{fig2} for signal and pile up background. It is clear that requesting this matching will lead to a negligible background~\cite{gammagamma5}. This conclusion remains for any exclusive production at the LHC, such as $WW$, $ZZ$, $t \bar{t}$, $\gamma Z$, etc.

\section{Quasi-exclusive dilepton production}

\begin{figure}
\centering
\includegraphics[width=0.6\textwidth]{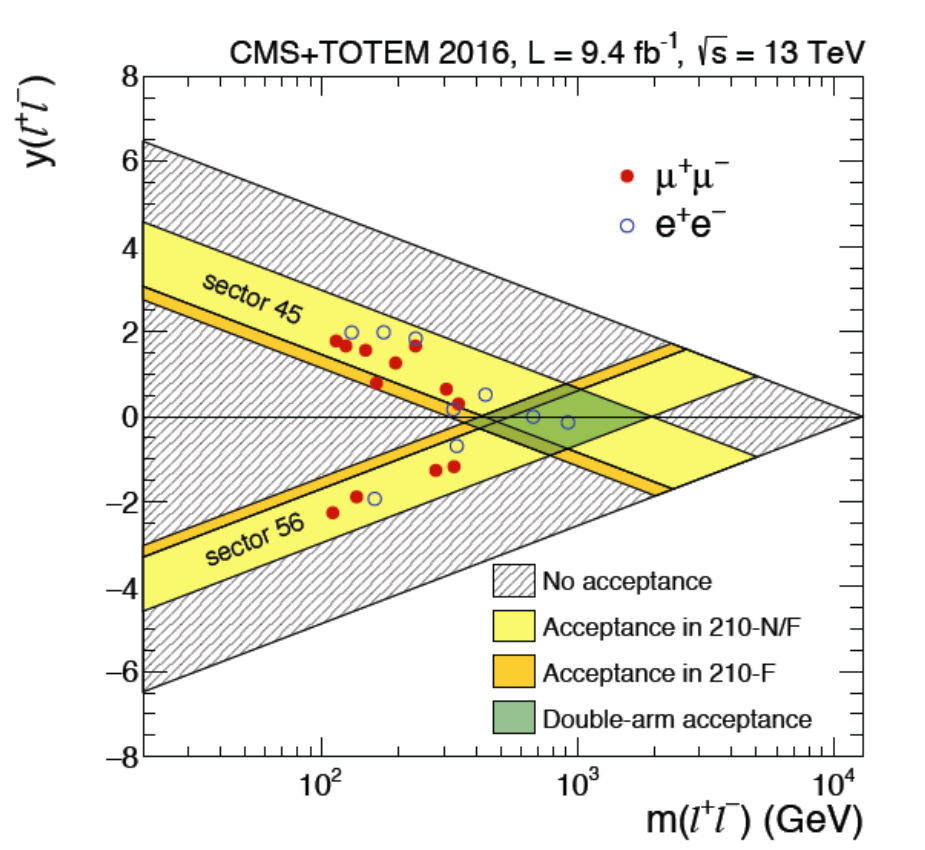}
\caption{Quasi-exclusive dilepton production. rapidity $y$ versus mass distribution for dilepton ($ee$ and $\mu \mu$) events.}
\label{fig3}
\end{figure}

In this section, we will describe the search for quasi-exclusive dilepton in PPS. Dilepton production is a QED process and we aim at detecting at least one intact proton in the final state (requesting both protons to be intact leads to less than 1 event since the cross section for dilepton masses above 450 GeV is very small). 
17 (resp. 23) events are found with protons in the PPS acceptance and 12 (resp. 8) show a
matching better than by 2$\sigma$ in the $\mu \mu$ (resp. $ee$) channel.
The significance is thus better than 5$\sigma$ for observing 20 events for a
background of $3.85$ ($1.49\pm 0.07 (stat) \pm 0.53 (syst)$ for $\mu \mu$ and 
$2.36\pm 0.09 (stat) \pm 0.47 (syst)$ for $ee$)~\cite{dilepton}. The rapidity versus mass of the dilepton system is shown in Fig.~\ref{fig3}  where we see the quasi-exclusive dimuons in red and dielectrons in empty circles.

\section{Exclusive production of diphotons}

In this section, we will present the recent results concerning the exclusive production of diphotons. The number of events predicted by the Standard Model (SM) is negligible in the acceptance of PPS. Extra-dimensions, composite Higgs models or axion-like particles for example can predict quartic $\gamma \gamma \gamma \gamma$ anomalous couplings and thus a much higher number of events that could be measured in PPS~\cite{gammagamma5}.
Anomalous couplings can appear via loops of new particles coupling to photons or via resonances decaying into two photons.
The search for exclusive diphoton production was performed by requesting back-to-back, high diphoton mass ($m_{\gamma \gamma}>350$ GeV), and a matching in rapidity and mass between diphoton and proton information.
The first limits on quartic photon anomalous couplings at high diphoton masses  were derived with about 10 fb$^{-1}$ of data ($|\zeta_1|<2.9~10^{-13}$ GeV$^{-4}$, $|\zeta_2|<6.~10^{-13}$ GeV$^{-4}$) with about 10 fb$^{-1}$~\cite{gammagamma8} and were updated with 102.7 fb$^{-1}$ ($|\zeta_1|<7.3~10^{-14}$ GeV$^{-4}$, $|\zeta_2|<1.5~10^{-13}$ GeV$^{-4}$)~\cite{gammagamma7}. 

\begin{figure}[h]
\centering
\includegraphics[width=0.5\textwidth]{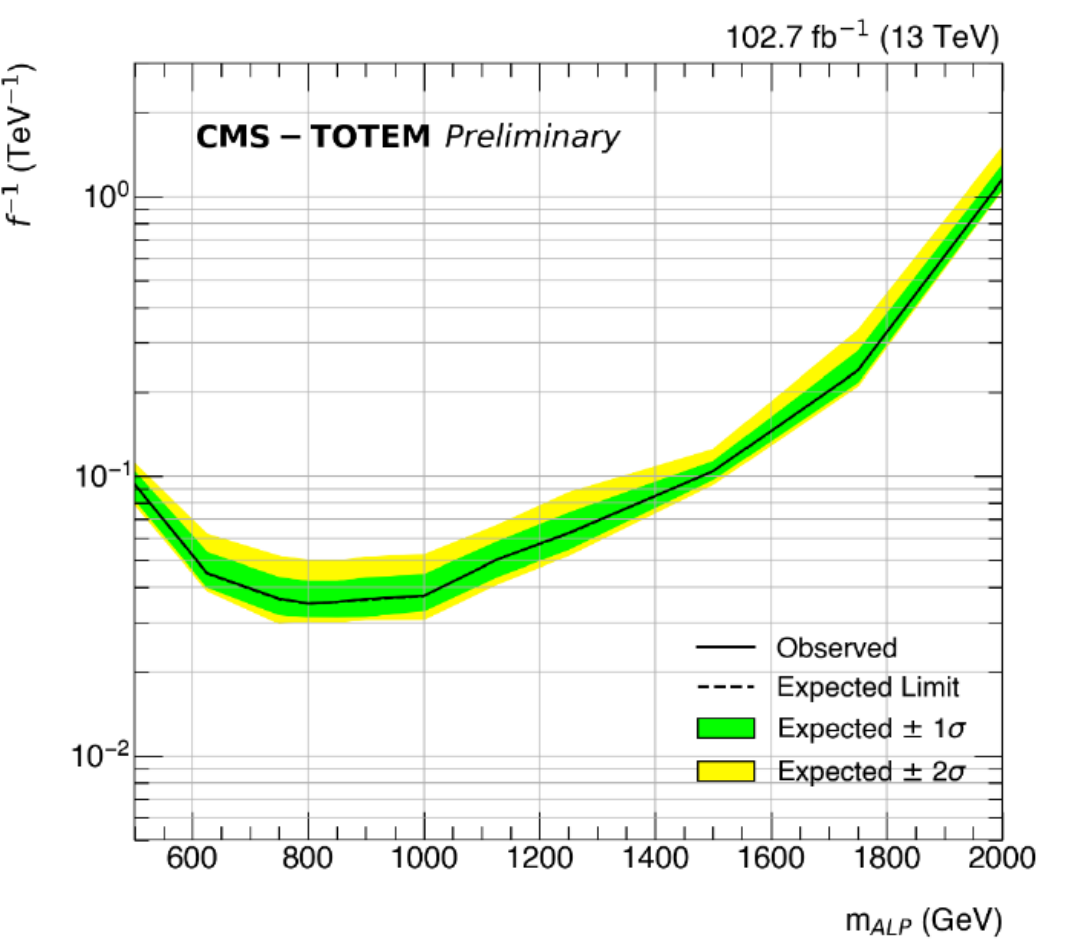}
\caption{First limits on the production of axion-like particles in the coupling versus mass plane using the exclusive production of two photons at the LHC.}
\label{fig4}
\end{figure}

The search for exclusive diphotons can be reinterpreted directly as a search for axion-like particles (ALP) decaying into two photons. The first limits on ALPs at high mass are shown in Fig.~\ref{fig4}~\cite{gammagamma7}.
The sensitivities were projected with 300 fb$^{-1}$ in Ref.~\cite{axion} using the same method and allow to gain about two orders of magnitude on sensitivity with respect to more standard methods at the LHC for ALP masses around 1 TeV and to cover in addition a domain at higher masses. It is also worth mentioning that this search for ALPs is complementary to the one performed at lower masses in heavy ion collisions~\cite{axion1}.

\section{Exclusive production of $W$ and $Z$ boson pairs}

Using the same method as for exclusive $\gamma \gamma$ production, it is possible to look for $WW$ and $ZZ$ exclusive production at the LHC. The search was performed in the full hadronic decay modes of the $W$ and $Z$ bosons since the anomalous production of $WW$ or $ZZ$ events dominates at higher mass with respect to the SM case, with a rather low cross section (the branching ratio of $W$ and $Z$ bosons decaying into hadrons is of course the highest one).  Two ``fat" jets  of radius 0.8 with  jet $p_T>200$ GeV, 1126$<m_{jj}<$2500 GeV, asking the jets to be back-to-back ($|1-\phi_{jj}/\pi|<0.01$) are requested. The fat jet correspond to the boosted decay of the $W$ or $Z$ boson. As usual, matching between the central $WW$ system and the proton information in rapidity and mass is also requested.

\begin{figure}[h]
\centering
\includegraphics[width=0.4\textwidth]{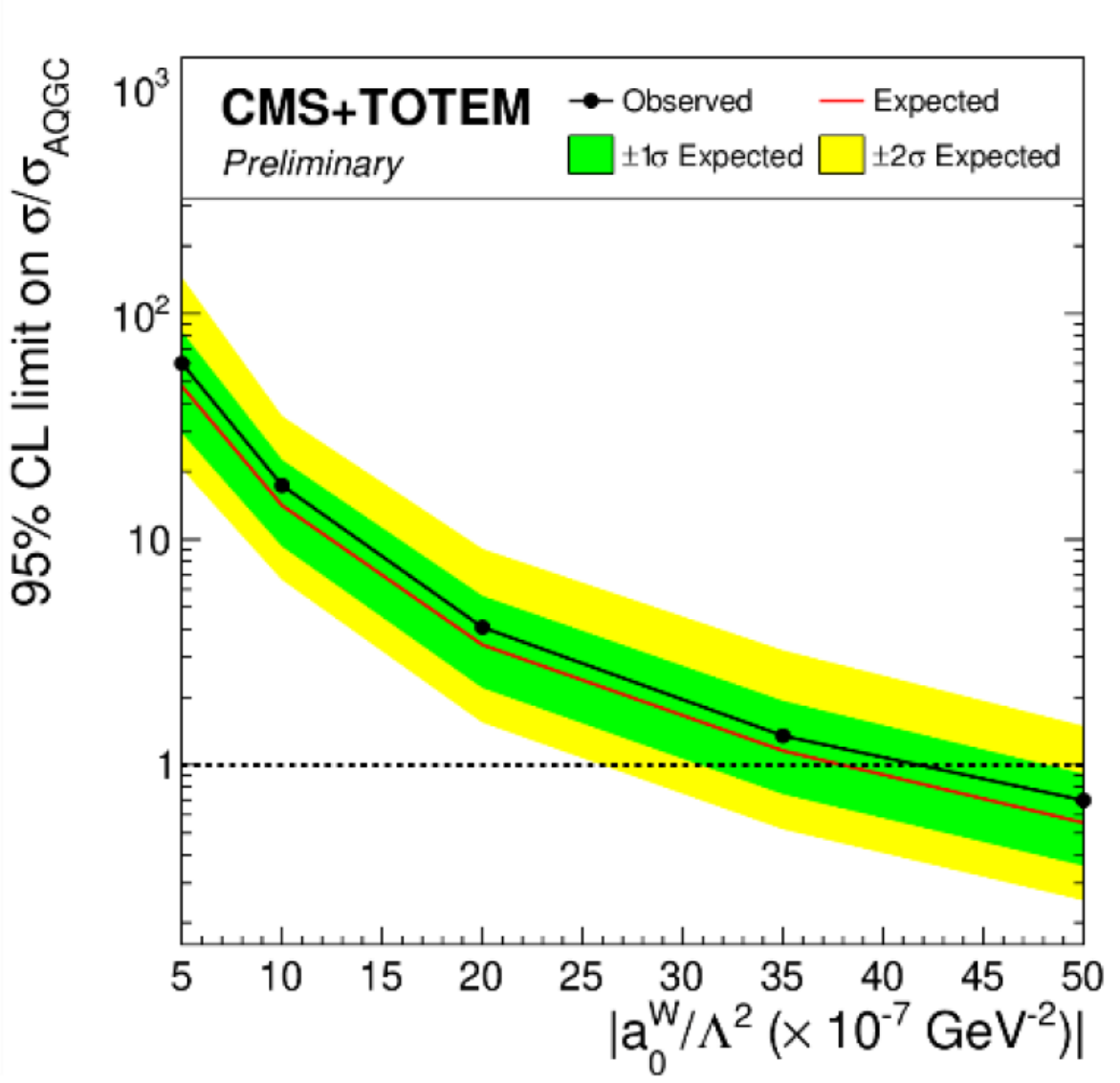}
\caption{Limits on quartic anomalous $\gamma \gamma WW$ $a_0^W$ coupling.}
\label{fig5}
\end{figure}

No signal was found and limits on the SM cross section were put as  $\sigma_{WW}<67$fb, $\sigma_{ZZ}<43$fb for $0.04<\xi<0.2$~\cite{ww1}. New limits on quartic anomalous couplings were determined as $a_0^W/\Lambda^2 < 4.3~10^{-6}$ GeV$^{-2}$, $a_C^W/\Lambda^2 < 1.6~10^{-5}$ GeV$^{-2}$, $a_0^Z/\Lambda^2 < 0.9~10^{-5}$ GeV$^{-2}$, $a_C^Z/\Lambda^2 < 4.~10^{-5}$ GeV$^{-2}$ with 52.9 fb$^{-1}$, and are shown in Fig.~\ref{fig5} for $a_0^W$ as an example.

The SM contribution of exclusive $WW$ production appears at lower $WW$ masses compared to anomalous couplings.
In order to observe this process at the LHC with higher luminosity (300 fb$^{-1}$ for instance), it is possible to use purely leptonic  channels for $W$ decays since the dijet background is too high at low masses for hadronic channels.
The SM prediction on exclusive WW (leptonic decays) after selection is about 50 events for 300 fb$^{-1}$  for 2 background events, which should lead to an observation of this process~\cite{ww}.

It is also worth noticing that it is possible to look in addition for the exclusive production of $\gamma Z$ events using the same method where the $Z$ boson decays either leptonically or hadronically. It leads to an improvement on the reach for the $\gamma \gamma \gamma Z$ anomalous coupling by about three orders of magnitude compared to the more standard method at the LHC which looks for the decay of the $Z$ boson into 3 photons~\cite{gammaz}.

\section{Exclusive production of $t \bar{t}$}

The search for exclusive $t \bar{t}$ production in leptonic and semi-leptonic decay modes was performed using about 29.4 fb$^{-1}$ of data. Because of the neutrino originating from the $W$ boson decay, the matching between the diproton and $t \bar{t}$ information is not enough to reject completely the background and kinematic fitters based on $W$ and $t$ mass constraints were used to reduce further the background.  No event was found and a limit was put on the $t \bar{t}$ exclusive production cross section as $\sigma^{excl.}_{t \bar{t}} < 0.59$ pb~\cite{ttbar1}.

It will be possible to improve significantly the sensitivities to $\gamma \gamma t \bar{t}$ anomalous couplings at the LHC by measuring the time of flight of the protons and requesting them to originate from the same vertex as the $t \bar{t}$ in order to reject the pile up background~\cite{ttbar}.

\section{Looking for $Z+X$ and $\gamma +X$ production}

\begin{figure}[h]
\centering
\includegraphics[width=0.8\textwidth]{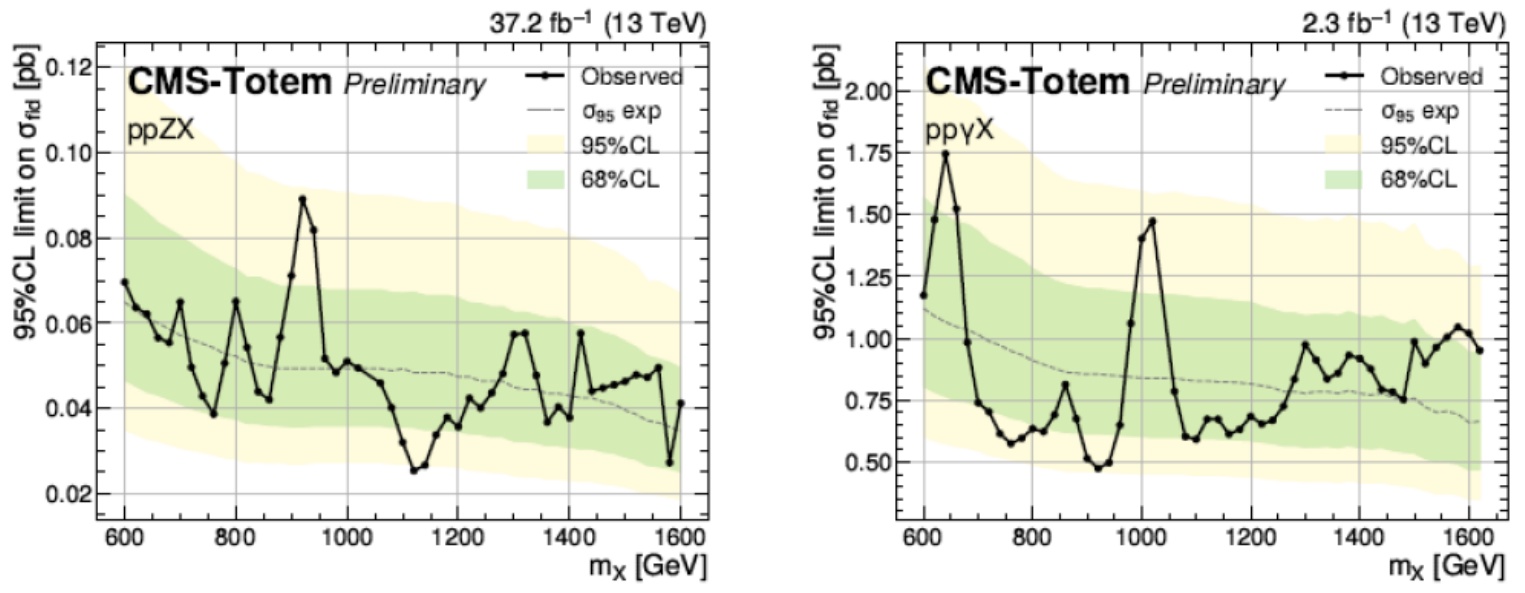}
\caption{95\% CL limit for $Z+X$ and $\gamma +X$ production as a function of $m_X$.}
\label{fig6}
\end{figure}

An additional original search for $Z+X$ and $\gamma +X$ events was also performed by the CMS and TOTEM collaboration. The idea to measure the total mass in the event reconstructed using the intact protons, allowing to obtain the mass of $Z+X$ and $\gamma + X$ while $X$ might  not be reconstructed. 
No signal was found and the limits are shown on Fig.~\ref{fig6}~\cite{zx}. This should be definitely redone with higher luminosity. 

To conclude, we described many results using PPS from the TOTEM and CMS collaborations, ranging from the observation of the quasi-exclusive production of leptons to the high sensitivity to quartic anomalous couplings such as $\gamma \gamma \gamma \gamma$, $\gamma \gamma WW$, $\gamma \gamma ZZ$, $\gamma \gamma t \bar{t}$ and to ALP. Better sensitivities is even expected for the next runs at the LHC using higher luminosities and improved detectors such as the fast timing ones.

\end{document}